\documentstyle[prl,aps,psfig,multicol]{revtex}
\begin{document}
\title{Metallic low-temperature resistivity in 2D over an extended
temperature range}
\author{S.~V.~Kravchenko}
\address{Physics Department, Northeastern University, Boston, Massachusetts
02115}
\author{T.~M.~Klapwijk}
\address{Department of Applied Physics, Delft University of Technology,
2628 CJ Delft, The Netherlands}
\date{\today}
\maketitle
\begin{abstract}
We report measurements of the zero-field resistivity in dilute 2D electron
system in silicon at temperatures down to 35~mK.  This extends the
previously explored range of temperatures in this system by almost an order
of magnitude.  On the metallic side, the resistivity near the
metal-insulator transition continues to decrease with decreasing
temperature and shows no low-temperature up-turn.  At the critical electron
density, the resistivity is found to be temperature-independent in the entire temperature range from 35~mK~to~1~K.
\end{abstract}
\pacs{PACS numbers: 71.30.+h, 73.40.Qv, 73.40.Hm}\vspace{-5mm}
\begin{multicols}{2}

Although the behavior suggestive of a zero magnetic field metal-insulator
transition (MIT) in two dimensions (2D) [1-11] 
was first reported five years ago, it still remains unexplained, with proposed driving mechanisms ranging from
contaminations in the oxide to superconductivity, and from percolation to a
non-Fermi liquid state (see, {\it e.g.}, Refs.[2-12]).  
One of the unusual features of dilute 2D systems is a dramatic drop of the resistivity on the metallic side of the transition.
This drop becomes apparent below a characteristic temperature
$T^*\sim\frac{1}{3}\,T_F$ where $T_F$ is the Fermi temperature.  $T^*$ is
rather high but approaches zero at the critical particle density.  The
relatively high value of $T^*$ prompted classical explanations of the
problem \cite{altshuler99,dassarma99,altshuler99a}.  If the mechanism behind the unusual behavior is indeed classical, then quantum localization \cite{abrahams79} should eventually take over at low enough temperatures, and in this case there will be no zero-temperature metal-insulator transition.  This possibility calls for experiments to be done at the lowest temperatures possible.

A decrease of the resistivity with decreasing temperature in 2D systems
(metallic behavior) does not by itself necessarily signal the presence of
any unanticipated physics.  It can be due, for example, to phonon
scattering, as it is in p-GaAs/AlGaAs heterostructures at relatively high
hole densities $\sim10^{11}$~cm$^{-2}$ \cite{mills99}.  At these densities,
the conductivity is $\gtrsim300\;e^2/h$, and the system may look metallic
down to inaccessibly low temperatures because logarithmic quantum
corrections to the conductivity are only $\sim e^2/h$ per decade in
temperature.  But at the particle density, $n_s$, close to the critical
density for the
metal-insulator transition, $n_c$, the conductivity becomes of the order of
$e^2/h$, and the decrease in temperature by a decade should now produce a
decrease in conductivity that is comparable to the conductivity itself.
Quantum corrections can only be seen at temperatures $T\ll T_F$.  In 
p-GaAs/AlGaAs heterostructures, $T_F$ at or near the critical hole density is very low, and the condition $T\ll T_F$ is hard to fulfill.  For example, in Ref.\cite{mills99}, $T_F$ is only about 300~mK\cite{mass} at the lowest (still metallic) hole density, and although temperatures as low as 40~mK have been reached in that paper, the lowest $T$ was of the order of $0.1\,T_F$.  More convenient systems to reach the condition $T\ll T_F$ at $n_s\sim n_c$ are silicon
metal-oxide-semiconductor field-effect transistors (MOSFET's) and p-SiGe
heterostructures, where $T_F$ at $n_s=n_c$ is of the order of 5 to 10
Kelvin.  However, although occasional data at temperatures as low as 16~mK
for Si~MOSFET's\cite{pudalov98} and 50~mK for p-SiGe\cite{coleridge97} were reported, the resistivity at electron/hole densities close to the MIT was carefully studied only down to $^3$He temperatures (250-300~mK) \cite{pudalov97,feng99,sarachik99}.

In this Letter, we report studies of resistivity in high-mobility Si
MOSFET's at temperatures down to 35~mK which extends the previously
explored temperature range by almost an order of magnitude.  We concentrate
on electron densities near the metal-insulator transition, where the
conductivity is $\lesssim e^2/h$.  We show that even at these very low
values of conductivity, but on the metallic side of the MIT ($n_s>n_c$),
the metallic temperature dependence of the resistivity, characterized by
$dR/dT>0$, persists down to the lowest accessed temperatures, which are less
than $10^{-2}\;T_F$ so the condition $T\ll T_F$ is well satisfied.  On the insulating side of the transition ($n_s<n_c$),
the behavior is always insulating, with $dR/dT<0$.  Between the metallic and insulating behaviors, at $n_s=n_c$, there exists a resistivity, $\rho_c$, which is practically temperature-independent in the entire temperature range from 35~mK to 1~K.  If one admits that the MIT-like behavior in two dimensions is merely a result of a competition between quantum localization and a classical
decrease of the Drude resistivity, then the existence of such a
temperature-independent curve would require cancellation of two
unrelated strong mechanisms in a wide temperature range, a coincidence
which seems hardly probable.

The samples used in our experiments were specially designed for
measurements at low electron densities and low temperatures by use of the
so-called split-gate technique.
\vbox{
\vspace{-12.5mm}
\hbox{
\hspace{0.50in}
\psfig{file=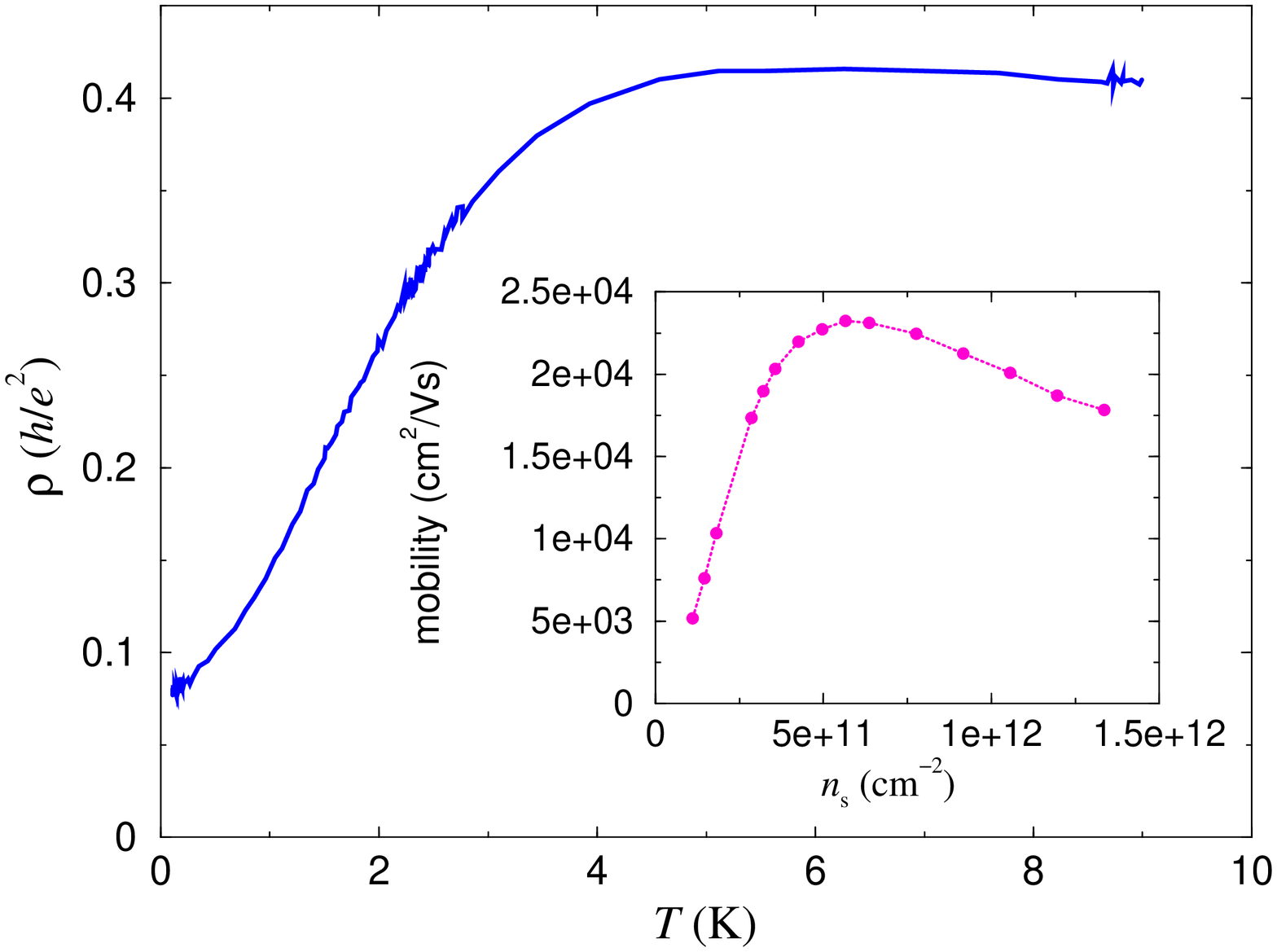,width=2.3in,bbllx=.5in,bblly=1.25in,bburx=7.25in,bbury=9.5in,angle=-90}
}
\vspace{0.3in}
\hbox{
\hspace{-0.15in}
\refstepcounter{figure}
\parbox[b]{3.4in}{\baselineskip=12pt \egtrm FIG.~\thefigure.
Resistivity as a function of temperature at
$n_s=1.11\times10^{11}$~cm$^{-2}$.  The inset shows electron mobility {\it vs}
electron density measured at 4.2~K.
\vspace{0.10in}
}
\label{fig1.ps}
}
}
Narrow ($<100$~nm) gaps in the gate metallization were
introduced near the doped contact areas which allowed for maintaining high electron density near the contacts and therefore dramatically reduced their resistance.  In the main part of the sample ($50\times120\,\mu$m$^2$) the electron density was controlled independently.  Typical mobility dependence of the electron density, $n_s$, is shown in the inset to Fig.~1.  The peak mobility of this sample was $2.3\times10^4$~cm$^2$/Vs at 4.2~K.  In the main part of the same figure, the temperature dependence of the resistivity is shown in the temperature range 100~mK to 9~K.  Experimentally, in the same material the higher the peak mobility of the electrons the larger the low-temperature drop of the resistivity \cite{kravchenko,popovic97,pudalov98}.  Indeed, in this sample the drop is somewhat smaller than that in the best Si~MOSFET's, being only a factor of 6 rather than $>10$.  However, the samples studied in this work were more convenient to use because of much higher quality of contacts which allowed reliable low-temperature transport measurements at these very low electron densities.

The samples were mounted at the end of a copper finger attached to the
mixing chamber of a Kelvinox-100 dilution refrigerator with base
temperature of 14~mK.  The electron gas was cooled by 16 thick (0.5~mm)
copper wires thermally coupled to the cold finger and mixing chamber, and
connected to the contacts on the sample holder.  Temperature was measured
by a ruthenium oxide thermometer placed near the sample and connected to
the mixing chamber and the cold finger by the same wires as the sample.
Temperature readings by the thermometer were typically 10~mK higher than
the temperature of the mixing chamber.  To reduce the external noise, three
sets of low-pass filters were used:  R-C-R filters with a cut-off frequency
of 40~Hz and two sets of C-L-C filters with a
cut-off frequency of 500~kHz.  Resistivity was measured by a four-terminal
{\it dc} method, while holding the
voltage drop across the whole sample (between two opposite current leads) to be less than 200~$\mu$V to ensure that the total power dissipated in the sample was less than $10^{-13}$~W.  The Fermi temperature was calculated using the electron density determined from the positions of the 
Shubnikov-de~Haas oscillations in a weak perpendicular
\vbox{
\vspace{-9mm}
\hbox{
\hspace{0.4in}
\psfig{file=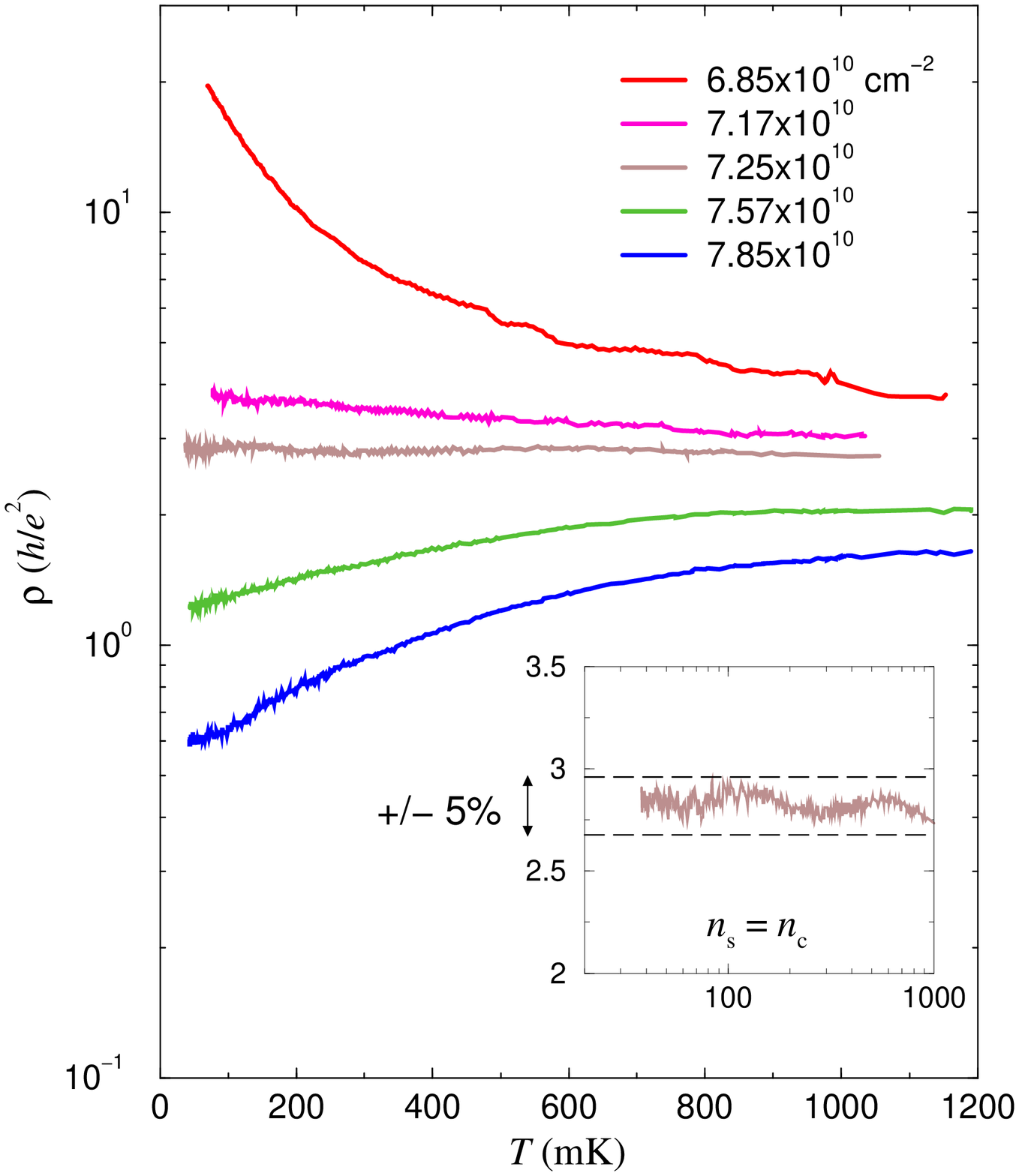,width=2.3in,bbllx=.5in,bblly=1.25in,bburx=7.25in,bbury=9.5in,angle=0}
}
\vspace{0.3in}
\hbox{
\hspace{-0.15in}
\refstepcounter{figure}
\parbox[b]{3.4in}{\baselineskip=12pt \egtrm FIG.~\thefigure.
Resistivity vs temperature at five electron densities as labeled.  The inset shows that the middle curve ($n_s=7.25\times10^{10}$~cm$^{-2}$) changes by less than $\pm5\%$ in the entire temperature range.
\vspace{0.10in}
}
\label{fig2.ps}
}
}
magnetic field and the value of the effective mass $0.19\,m_e$.  
The fact that the electron density does not change upon the application of the magnetic field was checked independently by
low-temperature measurements of the Hall voltage and capacitance.

Figure~2 shows the temperature dependence of the resistivity at five different electron densities at temperatures down to 35~mK.  The two upper curves display insulating behavior with $dR/dT<0$ and two lower curves display metallic behavior with $dR/dT>0$, in the entire temperature interval.  The middle curve shows almost no temperature dependence.  Deviations from the average value $\rho_c=2.82\,h/e^2$ do not exceed $\pm5\%$ (see the inset); at low temperatures, typically $\lesssim300$~mK, these deviations are not reproducible from one cool-down to another and even in different temperature sweeps.  Both above and below the temperature-independent curve the resistivity continues to change strongly down to the lowest temperature reached.  This confirms that the temperature of the electron gas continues to change down to the lowest temperatures.  Note that the change in the electron density, resulting in a sharp transition from strongly insulating to strongly metallic behavior, is less than 15\%.  In particular, a decrease of $n_s$ by only 1\% from its critical value is enough to cause a pronounced insulating behavior (see the second curve from the top which could be
mistakenly identified as a ``tilted separatrix'' if the tuning of the electron density were not fine enough).  Another obvious necessary condition to observe the flat separatrix is the high homogeneity of the electron density throughout the sample.

In the low-temperature limit, the two lower curves display metallic,
near-linear temperature dependence, as shown in Figs.~3~(a)~and~(b).  The resistivity shows
\vbox{
\vspace{-13mm}
\hbox{
\hspace{0.1in}
\psfig{file=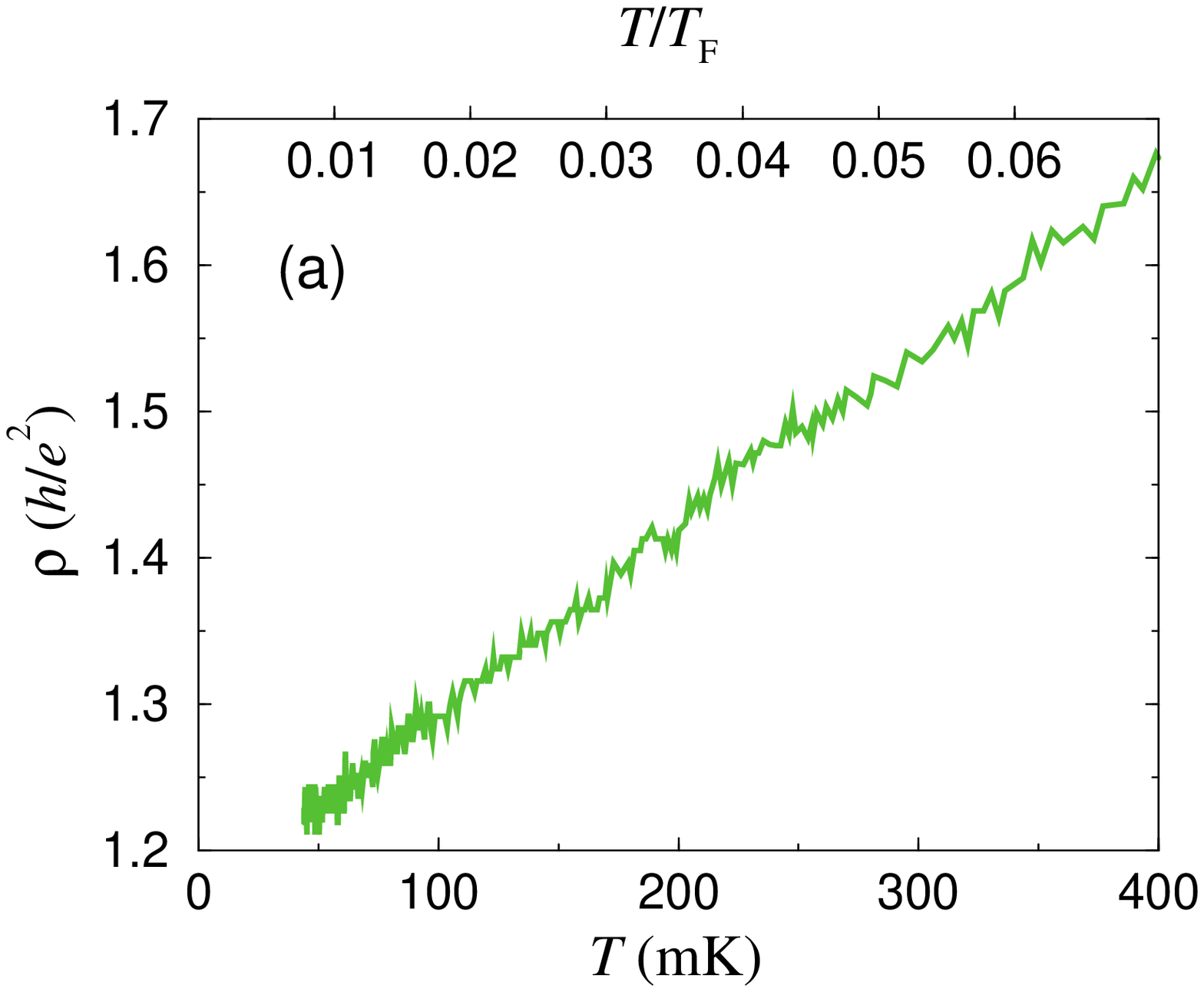,width=3in,bbllx=.5in,bblly=1.25in,bburx=7.25in,bbury=9.5in,angle=-90}
}}
\vbox{
\vspace{-10mm}
\hbox{
\hspace{0.1in}
\psfig{file=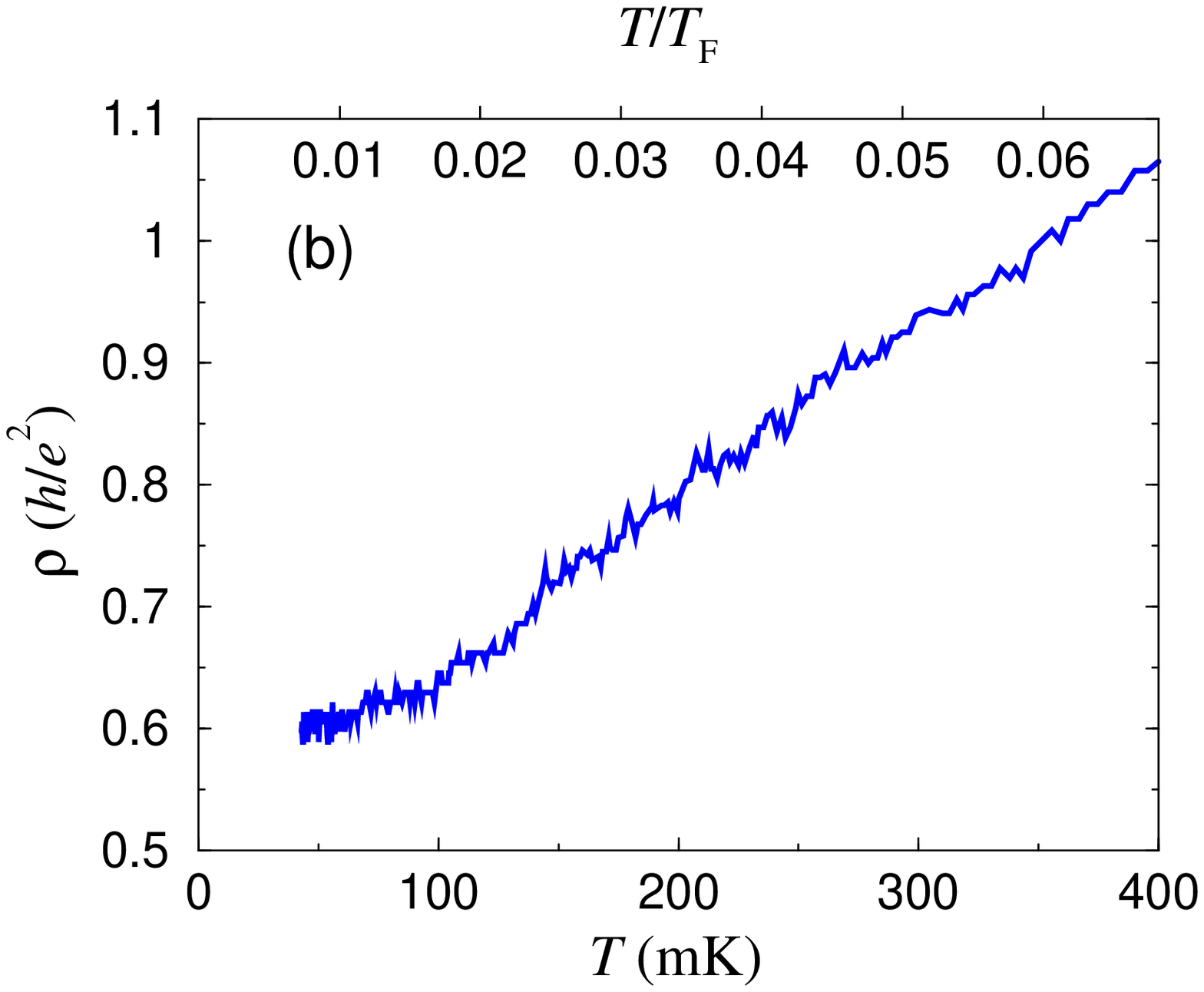,width=3in,bbllx=.5in,bblly=1.25in,bburx=7.25in,bbury=9.5in,angle=-90}
}
\hbox{
\hspace{-0.15in}
\refstepcounter{figure}
\parbox[b]{3.4in}{\baselineskip=12pt \egtrm FIG.~\thefigure.
Resistivities at $n_s=7.57\times10^{10}$~cm$^{-2}$ (a) and
$7.85\times10^{10}$~cm$^{-2}$ (b) as functions of temperature (lower
x-axes) and as functions of the ratio $T/T_F$ (upper x-axes).
\vspace{0.10in}
}
\label{fig3.ps}
}
}
no insulating up-turn in the low-temperature limit.  The lowest temperature reached corresponds to a ratio $T/T_F$ of less than $8\cdot10^{-3}$ (see upper x-axes in both figures).  The almost linear temperature dependence observed near the MIT at these low temperatures is different from the exponential temperature dependence seen at higher temperatures and higher densities \cite{pudalov97,hanein98,mills99}.

A striking feature of the data is the existence of a practically 
temperature-independent curve at $n_s=n_c$ which separates metallic
($dR/dT>0$) and insulating ($dR/dT<0$) behaviors.  At resistivity levels on
the order of or greater than $h/e^2$, one is in the regime where
$k_Fl\lesssim1$ (here $k_F$ is the Fermi vector and $l$ is the mean free
path).  To illustrate the expected strength of the quantum corrections in
this case, in Fig.~4 we plot the experimentally measured
$\rho(n_c)$ together with the resistivity calculated from the one-parameter
scaling theory (upper curve).  We assumed that the single parameter scaling
theory of localization holds even in the presence of very strong electron-electron interactions and that the Drude resistivity is
temperature-independent in this temperature range.  We use the standard formula \cite{abrahams79}
\begin{equation}
\frac{d\,\text{ln}\,\rho(L_\phi)}{d\,\text{ln}\,L_\phi}=-\beta(\rho).
\end{equation}
Here $L_\phi\propto\rho^{-\gamma}T^{-p/2}$ is the phase-breaking length,
$p$ and $\gamma$ are constants taken (after Ref.\cite{altshuler99a}) to be
equal to 3 and 0.5 respectively, and $\beta(\rho)$ is the scaling function \vbox{
\vspace{-4mm}
\hbox{
\hspace{0.40in}
\psfig{file=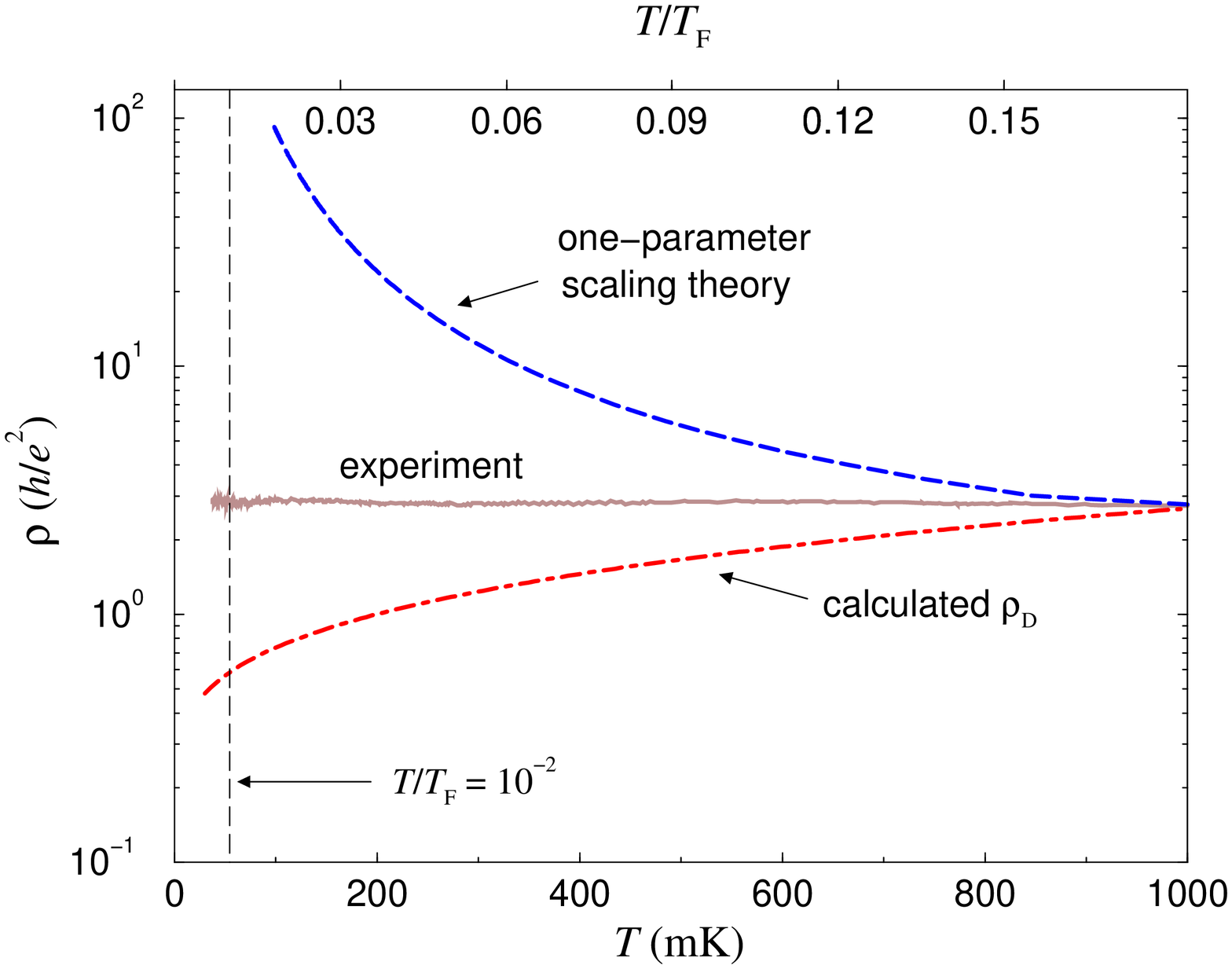,width=2.3in,bbllx=.5in,bblly=1.25in,bburx=7.25in,bbury=9.5in,angle=-90}
}
\vspace{0.3in}
\hbox{
\hspace{-0.15in}
\refstepcounter{figure}
\parbox[b]{3.4in}{\baselineskip=12pt \egtrm FIG.~\thefigure.
Resistivity at the separatrix (the middle curve of Fig.~2) as a function of temperature (lower x-axis) and as a function of the ratio $T/T_F$ (upper x-axis) compared to that calculated using one-parameter scaling theory (the dashed curve); see text.  As the dashed line shows, by the time the temperature reaches 100~mK, the resistivity of a ``conventional'' 2D system would have increased by a factor of more than 30.  The dot-dashed lower curve shows the calculated Drude resistivity required to produce the temperature-independent resistivity between metallic and insulating behaviors.
\vspace{0.10in}
}
\label{fig4.ps}
}
}
approximated as $\beta(\rho)=-\text{ln}(1+a\rho)$ where $a=2/\pi$ and
$\rho$ is measured in units of $h/e^2$.  The boundary condition is that at
$T=1$~K the calculated resistivity is equal to the measured one:
$\rho(1\text{ K})=2.82$\cite{corrections}.  As shown by the dashed curve in Fig.~4, by the time the temperature reaches 100~mK, quantum localization would increase the resistivity by a factor of more than 30.

The observation of a temperature-independent resistivity at $n_s=n_c$ over such a wide temperature range and at temperatures where the electron system
is clearly fully degenerate is certainly consistent with the existence of
the zero-temperature quantum phase transition \cite{sondhi97}.  In
combination with the results of Ref.\cite{sarachik99} where the
temperature-independent curve (with essentially the same value of resistivity as in this work) was observed in the temperature range
250~mK~--~1.8~K in another 2D system in silicon, we further allege that
there is no appreciable temperature dependence at $n_s=n_c$ in the
temperature range from 35~mK to 1.8~K, {\it i.e.}, for temperatures between approximately $6.5\cdot10^{-3}\,T_F$ in the present work to $0.25\,T_F$ in Ref.\cite{sarachik99}.  At higher temperatures, the resistivity at $n_s=n_c$ starts to decrease slowly with increasing temperature (see, {\it e.g.}, Refs.\cite{kravchenko,pudalov98}).  One possible cause of this behavior can be quantum-classical crossover which leads to insulating-like temperature dependence at temperatures comparable to or greater than $T_F$\cite{dassarma99}.

The existence of a temperature-independent curve between insulating and metallic behaviors was also reported in another 2D system, p-GaAs/AlGaAs heterostructure \cite{hanein98a}, at temperatures below 150~mK.  Above this temperature, the resistivity slowly decreases with increasing temperature similar to its behavior in our system at $T\gtrsim1.8$~K.  Taking into account the much lower value of $T_F\sim800$~mK\cite{mass} in the sample used in Ref.\cite{hanein98a} compared to our system, one may suggest that as in our case, the loss of full degeneracy causes this behavior.

In samples with lower mobility, the temperature-independent curve between metallic and insulating behaviors no longer exists, as can be seen, {\it e.g.}, in Fig.~4 of Ref.\cite{feng99}:  below a certain temperature $\sim1$~K, the curve which is nearly flat at higher $T$, turns ``insulating''.  In very disordered samples, the metallic behavior is not seen at all \cite{kravchenko,popovic97}.

Of course, the fact that we have observed no insulating-like upturn of the
resistivity in a wide, but still restricted, range of temperatures does not necessarily mean that the system will remain metallic at zero
temperature.  As has been already mentioned, if the MIT-like behavior
observed by us and others were a result of competition between Anderson
localization and a temperature-dependent Drude resistivity, $\rho_D(T)$,
the localization would eventually dominate at low temperatures.  However,
the existence of the temperature-independent curve between metallic and
insulating behaviors makes this possibility unlikely.  Indeed, it would
require the Drude resistivity to have a very special temperature dependence
(see Eq.6 in Ref.\cite{altshuler99a}):
\begin{equation}
\frac{d\,\text{ln}\,\rho_D(T)}{d\,\text{ln}\,T}=\frac{p}{2}\left(1-\frac{1}{\beta(\rho_D(T))}\right)^{-1}.
\end{equation}
The Drude resistivity calculated using this equation with the same boundary
conditions as before is shown in Fig.~4 by the dot-dashed
line.  We note that the shape of $\rho_D(T)$ required to produce a
temperature-independent curve is determined by the $\beta$-function
which comes from the scaling theory, and the parameter $p$ which comes
from the relation between $L_\phi$ and $T$.  It would therefore be a
remarkable coincidence if the required $\rho_D(T)$ dependence over a
wide temperature range were produced by some classical mechanism not
related apriori to either $\beta$ or $p$ \cite{generality}.

In summary, we have shown experimentally that the strong metallic temperature
dependence of the resistivity in silicon MOSFET's survives down to at least
35~mK.  This extends the previously explored range of temperatures by
almost an order of magnitude.  At these temperatures, the system is clearly
fully degenerate with the ratio $T/T_F$ being less than $10^{-2}$.  At the
critical electron density, the resistivity is temperature-independent in
the entire temperature interval.  These observations put valuable
constraints on possible theoretical explanations.

We are grateful to S.~Bakker and R.~Heemskerk for their contributions in
developing and preparing the MOSFET's used in this work, to C.~Chamon,
D.~Heiman, V.~Ya.~Kravchenko, M.~P.~Sarachik, and D.~Simonian for useful
discussions, and to T.~Hussey and Y.~Zhu for technical assistance.  This
work was supported by the NSF Grant No.~DMR-9803440.\vspace{-5mm}


\begin{thebibliography}{10}
\bibitem{corrections} By choosing this boundary condition we assume that
the cut-off temperature at which localization effects become negligible is
of the order of 1~K.  In reality it is likely to be higher, and this would
lead to even stronger quantum corrections.
\bibitem{generality} We stress that Eq.(6) in Ref.[20] is obtained for a
very general case, without assuming any particular mechanism for the
possible temperature dependence of $\rho_D$.
\end{thebibliography}


\end{multicols}

\end{document}